\documentclass[5p]{elsarticle}

\journal{Nucl. Instrum. and Meth. in Phys. Res. A}

\usepackage{amsmath}
\usepackage{booktabs}
\usepackage{hyperref}
\usepackage{lineno}
\usepackage{textcomp}

\modulolinenumbers[5]

\biboptions{sort&compress}
\graphicspath{{figures/}} % Directory in which figures are stored

\newcommand{\milli}{m}
\newcommand{\micro}{\textmu}

\newcommand{\meter}{m}

\newcommand{\volt}{V}
\newcommand{\second}{s}
\newcommand{\watt}{W}

\newcommand{\celsius}{$^\circ$C}
\newcommand{\percent}{\%}

\renewcommand{\neq}{$n_{\text{eq}}\text{cm}^{-2}$}

\newcommand{\num}[1]{$#1$}

\newcommand{\SI}[2]{{#1}\,{#2}} 

\makeatletter
\def\@author#1{\g@addto@macro\elsauthors{\normalsize%
    \def\baselinestretch{1}%
    \upshape\authorsep#1\unskip\textsuperscript{%
      \ifx\@fnmark\@empty\else\unskip\sep\@fnmark\let\sep=,\fi
      \ifx\@corref\@empty\else\unskip\sep\@corref\let\sep=,\fi
      }%
    \def\authorsep{\unskip,\space}%
    \global\let\@fnmark\@empty
    \global\let\@corref\@empty  %% Added
    \global\let\sep\@empty}%
    \@eadauthor={#1}
}
\makeatother

\begin{document}

\begin{frontmatter}

\title{Investigation of modified ATLAS pixel implantations after irradiation with neutrons}

%\title{Elsevier \LaTeX\ template\tnoteref{mytitlenote}}

\author{A. Gisen\corref{cor1}}
\author{S. Altenheiner}
\author{C. G\"o\ss{}ling}
\author{M. Grothe}
\author{R. Klingenberg\fnref{fn1}}
\author{K. Kr\"oninger}
\author{J. L\"onker}
\author{M. Weers}
\author{T. Wittig\fnref{fn2}}
\author{F. Wizemann}

\address{TU Dortmund, Experimentelle Physik IV, 44221 Dortmund, Germany}
\cortext[cor1]{corresponding author, andreas.gisen@tu-dortmund.de}
\fntext[fn1]{deceased 24 May 2017}
\fntext[fn2]{now at CiS Forschungsinstitut f\"ur Mikrosensorik GmbH, Erfurt}

\begin{abstract}
The innermost part of the tracking detector of the ATLAS experiment consists mainly of planar n$^+$-in-n silicon pixel sensors. 
During the phase-0 upgrade, the Insertable B-Layer (IBL) was installed closest to the beam pipe.
Its pixels are arranged with a pitch of $250\,\mu$m$\,\times\,50\,\mu$m with a rectangular shaped n$^+$ implantation.
Based on this design modified pixel designs have been developed in Dortmund.

Six of these new pixel designs are arranged in structures of ten columns and were placed beside structures with the standard design on one sensor. Because of a special guard ring design, each structure can be powered and investigated separately. Several of these sensors were bump bonded to FE-I4 read-out chips. One of these modules was irradiated with reactor neutrons up to a fluence of $5 \times 10^{15} \, n_{\text{eq}}\text{cm}^{-2}$.

This contribution presents important sensor characteristics, charge collection determined with radioactive sources and hit efficiency measurements, performed in laboratory and test beam, of this irradiated device. It is shown that the new modified designs perform similar or better than the IBL standard design in terms of charge collection and tracking efficiency, at the cost of a slightly increased leakage current.
\end{abstract}


\end{frontmatter}

%\linenumbers

\section{Introduction}

One of the major experiments at the CERN LHC is ATLAS \cite{ATLAS:Experiment}. 
During the phase-0 upgrade in 2014, the Insertable B-Layer (IBL) \cite{ATLAS:IBL_TDR,ATLAS:IBL_TDR2} was installed closest to the beam pipe. 
As the other layers of the pixel detector, it consists of planar n$^+$-in-n silicon pixel sensors, but their design layout was revised \cite{1748-0221-7-02-C02051} and new front-end electronics, the FE-I4 read-out chips \cite{GARCIASCIVERES2011S155}, were used.
The IBL is designed to withstand a fluence of \SI{$5 \times 10^{15}$}{\neq} and a dose of \SI{2.5}{MGy} during its operation. 

In order to achieve a high particle tracking efficiency even after this irradiation, it is necessary to have a sufficient signal charge induced in the sensors.
For this purpose, bias voltages up to \SI{1000}{\volt} can be applied.
These higher operation voltages lead to higher power consumption, which has to be dissipated by the cooling system to prevent thermal runaway.
Therefore, it is desirable to achieve higher signal charge and efficiency at the same voltages or to achieve the same signal charge and efficiency at lower voltages.

Previous investigations showed an increased signal charge after irradiation for thinner detectors \cite{Casse2010} and at higher voltages \cite{Kramberger2010}, caused by charge multiplication.
The presented approach is another:
Modifications are made to the shape of the n$^+$ pixel implantations which are intended to force maxima in the electric field to increase the collected charge.

\section{Design of the pixel cells}
\label{sec:design}

\begin{figure}[htbp]
    \centering % \begin{center}/\end{center} takes some additional vertical space
    \includegraphics[width=0.9\linewidth]{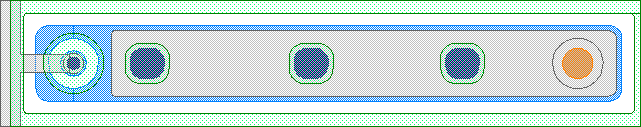}
    \caption{\label{fig:IBL_pixel_cell} Schematic layout of the IBL pixel design. The n$^+$ implantation is displayed in blue, the metal in grey, the nitride openings, which are also used for the moderated p-spray technique, in green and the opening in the outer passivation in orange.}
\end{figure}

The pixel cells of the IBL sensors have a pitch of \SI{250}{\micro\meter} in the long and \SI{50}{\micro\meter} in the short direction. Their design is shown in \autoref{fig:IBL_pixel_cell}.
The rectangular n$^+$ implantation is positioned centrally. Its corners are rounded in order to avoid maxima in electric field strength.
It is surrounded by openings in the nitride layer, which are used for the moderated p-spray technique.
The bias dot with the connection to the bias grid is on the left.
The opening in the outer passivation layer for the subsequent bump bond deposition is on the right.
In between there are three openings in the inner passivations, which ensure a conductive connection from the n$^+$ implantation through the metal layer to the bump bond.

\begin{figure}[htbp]
    \centering % \begin{center}/\end{center} takes some additional vertical space
    \includegraphics[width=0.9\linewidth]{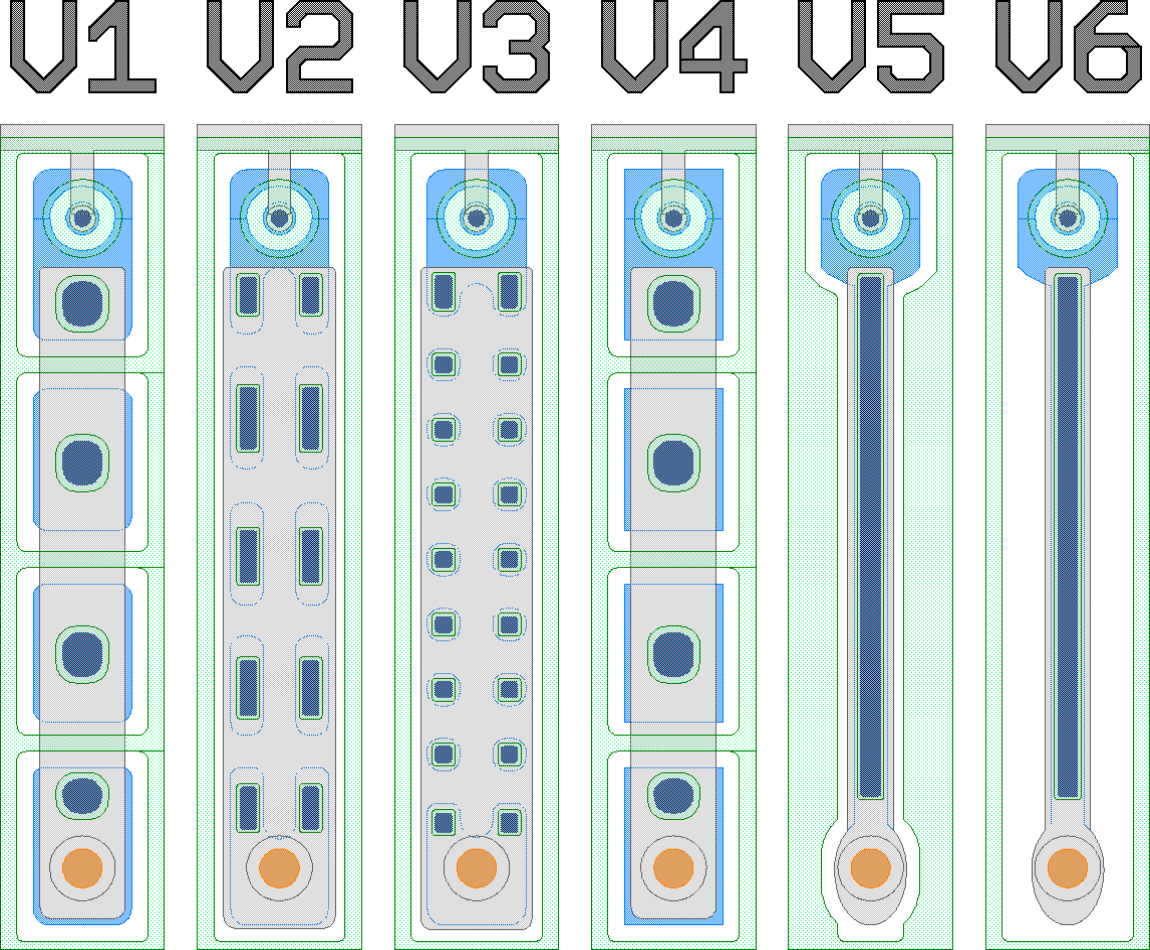}
    \caption{\label{fig:R_MkI_designs} Schematic layout of the modified designs V1 to V6. Again the n$^+$ implantation is displayed in blue, the metal in grey, the nitride openings, which are also used for the moderated p-spray technique, in green and the opening in the outer passivation in orange.}
\end{figure}

Based on this IBL pixel design, six modified so-called \emph{REINER}\footnote{\textbf{RE}designed, \textbf{IN}novative, \textbf{E}xciting and \textbf{R}ecognizable} pixel designs were developed in Dortmund \cite{Wittig2013}. They are numbered V1 to V6 and their layouts are shown in \autoref{fig:R_MkI_designs}.
In V1, the n$^+$ implantation is divided into four segments. The moderated p-spray profile is continued between the individual segments.
V2 and V3 feature further subdivisions to ten and 16 segments, respectively. Due to reduced space between the single segments, no moderated p-spray profile could be implemented.
V4 has a similar layout as V1, but with sharp corners of the segments. The usual rounding of the corners was dispensed.
V5 and V6 have the n$^+$ implantation narrowed by a factor of 3. 
V6 has the same moderated p-spray profile as the standard design, while this is changed in V5.
The width of the high dose area is largely increased.

All designs have an identical bias dot and bias grid structure. It is known \cite{Weingarten12,ATLAS:IBL_Prototype_modules} that these lead to reduced charge collection and hit efficiency, but improvements in this area are not part of this study.

\section{Sensors and modules}
\begin{figure}[htbp]
    \centering % \begin{center}/\end{center} takes some additional vertical space
    \includegraphics[width=0.95\linewidth]{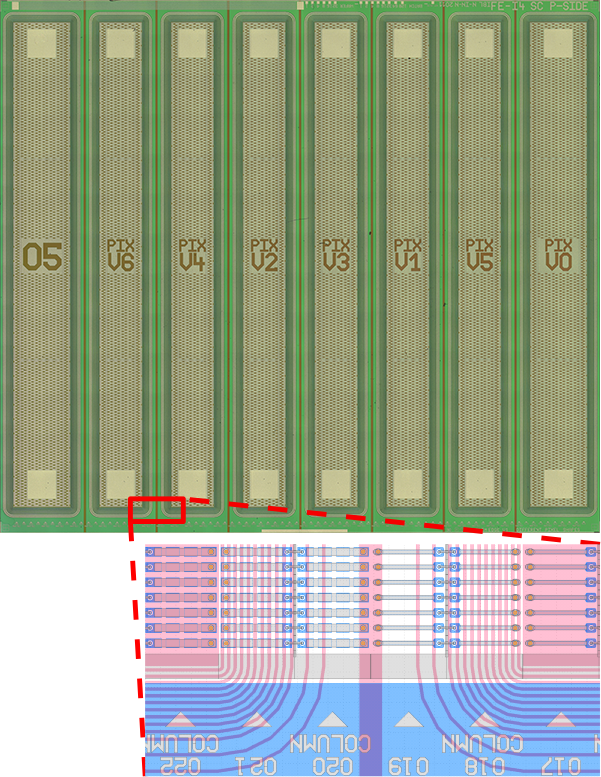}
    \caption{\label{fig:p-Side} Image of the p-side of a REINER pixel sensor. A detailed view on the guard ring layout between two designs is given.}
\end{figure}

The pixel designs presented in \autoref{sec:design} are placed on planar n$^+$-in-n silicon sensors with an n-bulk thickness of \SI{200}{\micro\meter}.
%These are processed on oxygenated n-doped float zone silicon wafers which carry on the front the highly n-doped pixel implantations with moderated p-spray isolation and a p-doped backside with guard rings.
%The n-bulk thickness is \SI{200}{\micro\meter}.
They are compatible with IBL single chip sensors. The pixel cells are arranged in 80 columns and 336 rows.
Each ten columns the pixel design is changed. 
While the pixels on the n-side are continuous, the p-side is divided by guard ring design into eight substructures.
Each of those features only one pixel design whose name is printed centrally.
High-voltage pads for contacting are located at the top and bottom.
Each structure is enclosed by 13 guard rings. This is pictured in an image of the p-side in \autoref{fig:p-Side}.	
The six structures containing the modified designs are enclosed by structures with the IBL design. Since this baseline design occurs twice, redundancy and comparability is given within a sensor.
The first structure with the IBL design is named 05, the sensor's number on the wafer, while the second appearance is named V0. The other structures have their corresponding structure name imprinted.

A detailed view of the guard ring layout at the bottom of \autoref{fig:p-Side} reveals that not all pixel cells are covered by the high-voltage electrode. The next-to-last column is fully shifted beneath the guard rings, the last column even beyond.
%The external dimensions of the active area are \SI{$20.5 \times 16.8$}{\milli\meter$^2$}

A few sensors were connected to FE-I4 \cite{GARCIASCIVERES2011S155} read-out chips with the help of a flip-chip process employing tin-lead-bumps.
This flip-chip process was performed at IZM, Berlin, Germany, and is consistent with IBL production.
The investigated module was irradiated with neutrons to \SI{$5 \times 10^{15}$}{\neq} at the Sandia Annular Core Research Reactor, Albuquerque, USA \cite{acrr_web}.
%\footnote{\url{http://www.sandia.gov/research/facilities/annular_core_research_reactor.html}}
Afterwards, it was mounted to a read-out PCB, which provides connection via wire bonds to the FE as well as connectors for low/high voltage supply and data transmission.

\section{IV and power dissipation measurements}
\label{sec:iv}
Measurements took place in an isolation box which includes a metal heat exchanger connected to a regulated chiller. With this setup, on-sensor temperatures down to \SI{-50}{\celsius} can be reached. The box is flushed with pre-cooled dried air to keep humidity low and therefore preclude condensation.
The FE was not powered during this study to prevent its heat load being transmitted to the sensor. 
Pt1000 temperature resistors were placed as close as reasonably possible to the module.
All measurements were taken at an on-sensor temperature of \SI{-30}{\celsius}, so scaling of the leakage current could be avoided. 

\begin{figure}[htbp]
    \centering % \begin{center}/\end{center} takes some additional vertical space
    \includegraphics[width=\linewidth]{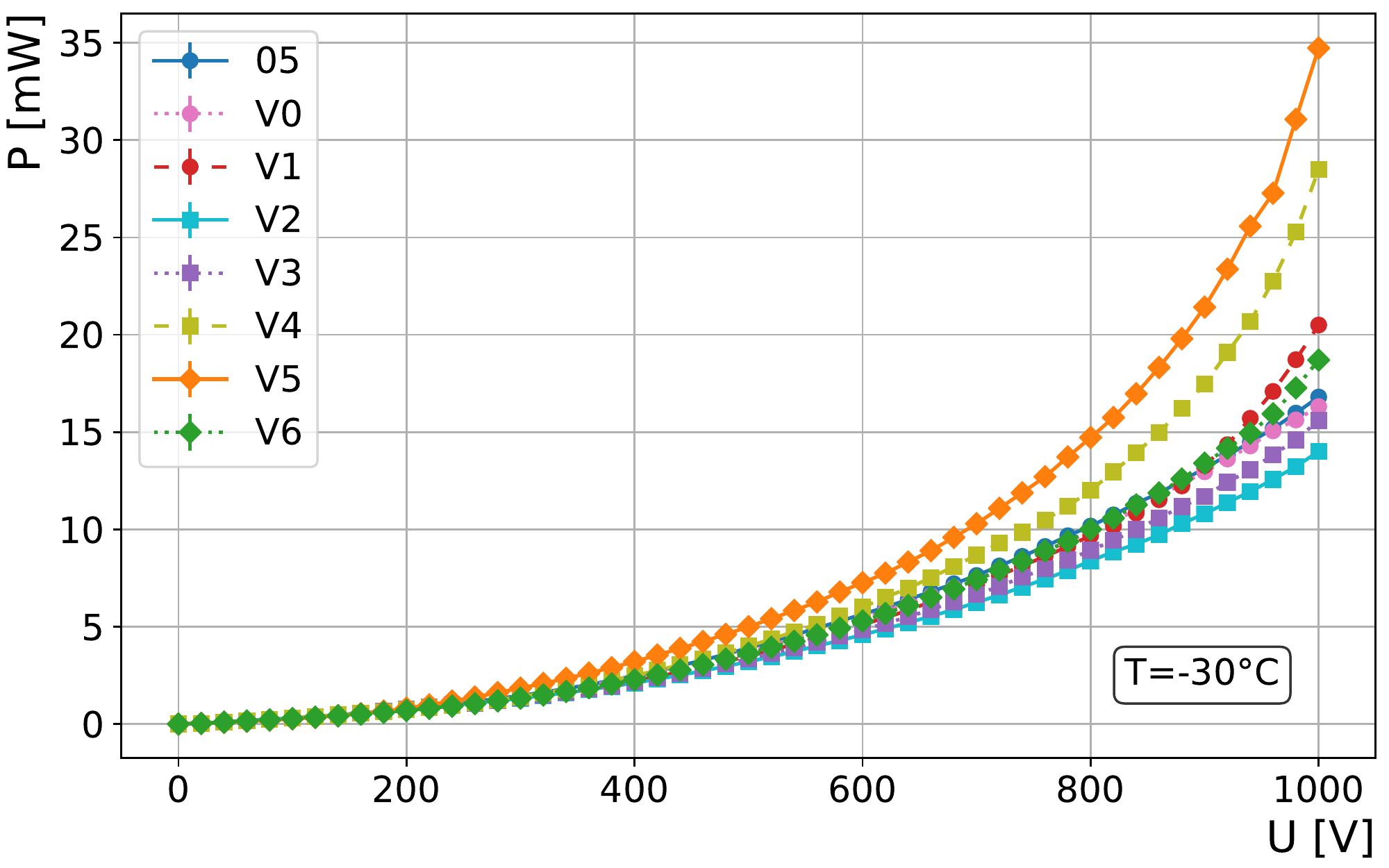}
    \caption{\label{fig:IV_single} Power dissipation of individual structures, at \SI{-30}{\celsius}.}
\end{figure}

Jumpers on the read-out PCB enable biasing of one, all or any combinations of structures during measurements.
The current was measured up to a maximum voltage of \SI{1000}{V} in steps of \SI{10}{\volt} with a delay of \SI{10}{\second} between the steps. At each step the current was measured ten times and the mean value was calculated. Differences between individual structures are visible, but none is going into break down.
The resulting power dissipation curves for the measurements of all structures individually are shown in \autoref{fig:IV_single}.
An indication for the reliability of the measurements is the good agreement of the curves of 05 and V0, the structures with the IBL design.
%Good agreement between the currents of 05 and V0, the structures with the IBL design, can be observed. 
Their mean deviation is \SI{$(0.133 \pm 0.001)$}{\milli\watt}, while their maximum deviation is \SI{0.49}{\milli\watt}.

A stronger rise can be observed for V5 and V4 from \SI{200}{\volt} upwards, and for V1 and V6 from \SI{900}{\volt} upwards, leading to increased power in comparison to the other designs. 
The clear differences between the quite similar designs, V5/V6 and V1/V4, are remarkable.
Especially at high voltages self-heating of the sensor cannot be excluded. However, since neither a significant increase in temperature nor a strong rise in each design is measured, it is very likely that the differences are caused by the different designs. 

\begin{figure}[htbp]
    \centering % \begin{center}/\end{center} takes some additional vertical space
    \includegraphics[width=\linewidth]{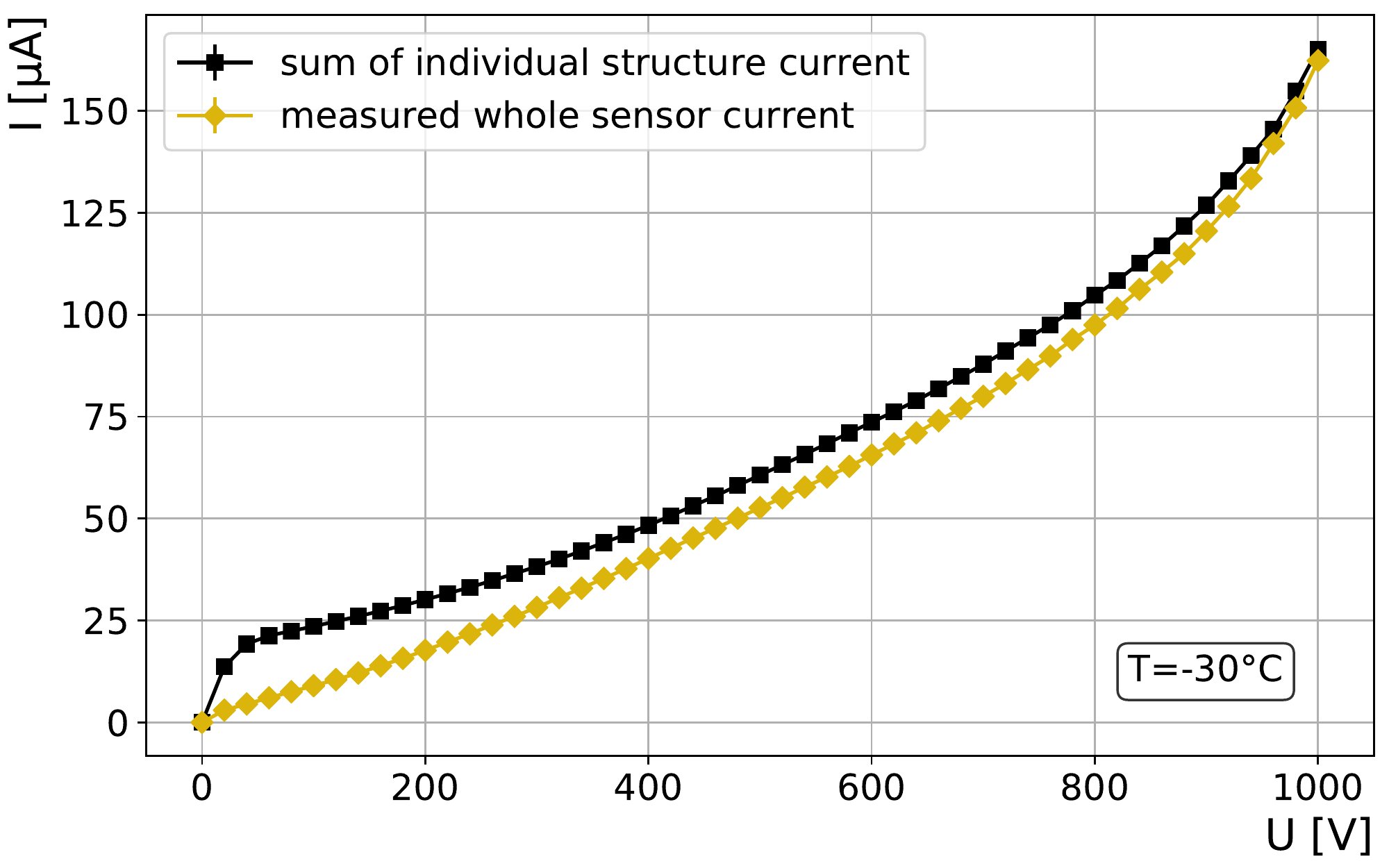}
    \caption{\label{fig:IV_sensor} IV comparison of individual structures being biased and the sensor as a whole.}
\end{figure}

%The currents of the individual measurements are summed and plotted along a measurement of the whole sensor, where all structures were biased at the same time.
The currents of the individual measurements are summed and plotted along with a measurement of the whole sensor in \autoref{fig:IV_sensor}.
All structures were biased at the same time for the latter measurement. 
Contrary to the naive expectation, a big difference between both curves is visible. The observed difference decreases for increasing voltage.
This effect was also observed at unirradiated REINER sensors and other n$^+$-in-n structures \cite{KLINGENBERG2016}, indicating that the leakage current contribution from lateral and edge effects is not negligible, especially for small area devices.
This should always be considered when statements of test structures are related to full-scale devices.

\section{Charge collection measurements}
\begin{figure}[htbp]
    \centering % \begin{center}/\end{center} takes some additional vertical space
    \includegraphics[width=0.9\linewidth]{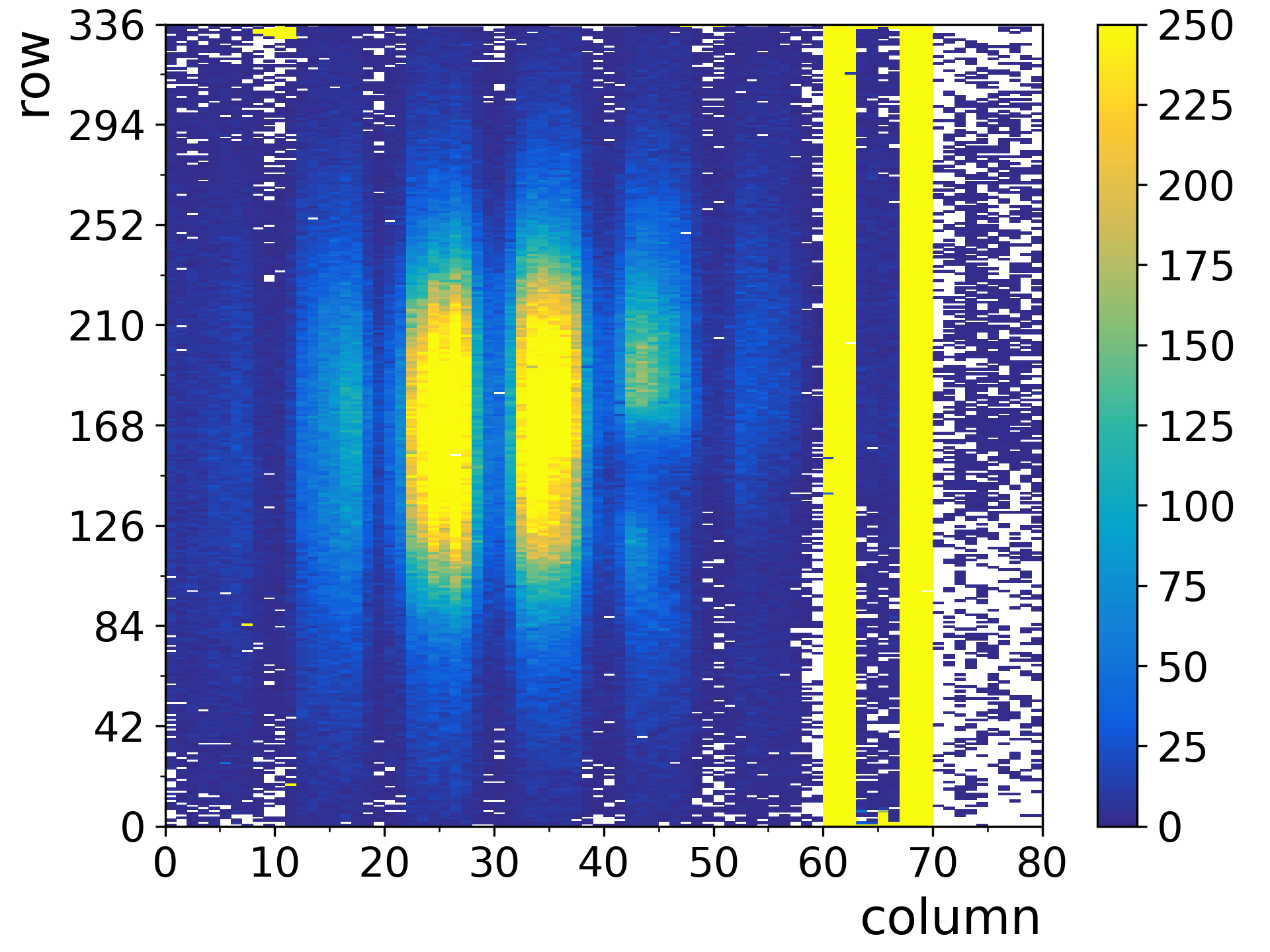}
    \caption{\label{fig:hitmap} Source scan hit map of the module, at \SI{600}{V}. All pixels were enabled during this scan.}
\end{figure}

A Strontium-90 beta source and a trigger scintillator were added on top of each other to the cooling setup described in \autoref{sec:iv}. 
The module in between was moved in such a way that the structure to be examined was always centered and aligned to the source setup.
% to use it for charge collection measurements.
%The same cooling setup as described in \autoref{sec:iv} was also used for source measurements, performed with an Sr-90 $\beta$-source.
The emitted electrons pass through the sensor and its read-out chip, a thin aluminium plate and finally the scintillator, which then sends a read-out trigger to the chip.
%Starting with this raw data, \textit{fei4Analyzer}\footnote{\url{https://github.com/terzo/fei4Analyzer}} is used to match hits in adjacent pixels to a hit cluster.		
A hit map of a source scan at \SI{600}{\volt} is shown in \autoref{fig:hitmap}. 
The beam spot is clearly visible, along with the inefficiencies caused by the guard rings between different structures.
Noticeable are the two noisy areas between columns 60 and 70, which are related to the guard rings of design V5.
%With increasing bias voltage, other structures are affected as well. It seems to be a current related issue.
With increasing bias voltage, the same effect can be observed at other structures.
Tests at different temperatures reveal a dependence. 
Since the leakage current is highly temperature-dependent and since design V5, which has the highest leakage current of all designs, is influenced at lower bias voltages, it seems to be a current-related issue.
The successful operation of pixels being shifted under guard rings has been shown several times \cite{1748-0221-7-02-C02051,Wittig2013,Gisen17}. Thus, this behaviour was not expected and seems to be an exclusive feature of this special sensor design with multiple long but narrow guard ring structures placed on the same sensor, also featuring some pixel columns being shifted beyond the guard rings.
As an outcome, read-out of all pixel columns in the area of the guard rings was disabled and they are therefore excluded in all further measurements. 
%After that, no further difficulties occurred.

Each pixel read-out cell of the FE-I4 chip contains a discriminator with an adjustable threshold.
If a signal in the sensor exceeds this threshold, the time over threshold (ToT) is measured.
Therefore the ToT is directly related to the charge induced in the sensor.
For all measurements, the FE was tuned to a threshold of \SI{3200}{e} and a ToT-response of \num{6} at a reference charge of \SI{20}{ke}. 
The on-sensor temperature was \SI{-25}{\celsius}.

\begin{figure}[htbp]
    \centering % \begin{center}/\end{center} takes some additional vertical space
    \includegraphics[width=1.\linewidth]{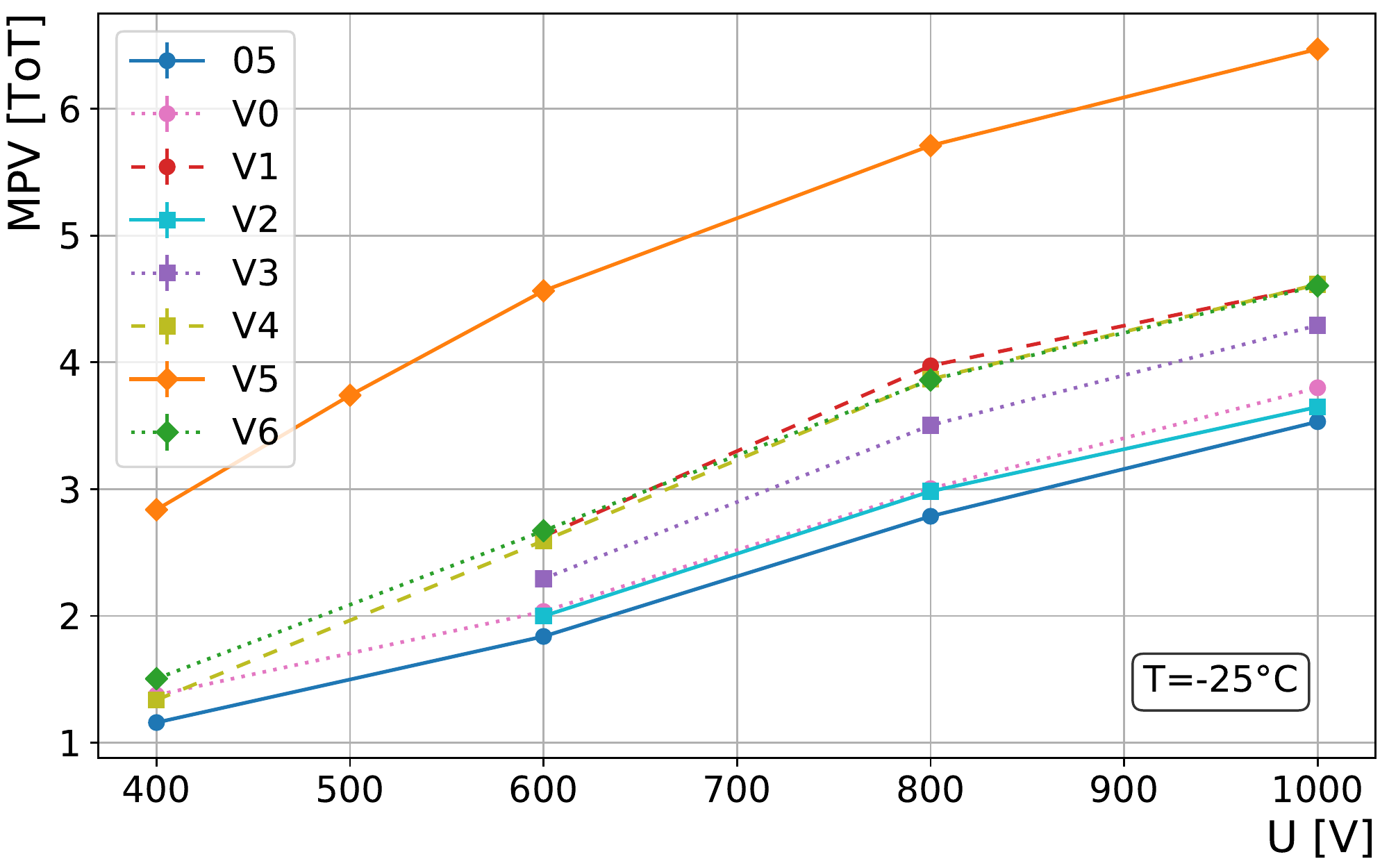}
    \caption{\label{fig:mpv_vs_bias} MPV vs. bias voltage, measured at a threshold of \SI{3200}{e} and a ToT-response of \num{6} at a reference charge of \SI{20}{ke}. The readout was enabled only for the examined structure. The error bars resulting from the fit are too small to be visible.}
\end{figure}

Measurements were done for all structures at several bias voltage points ranging from \SI{400}{\volt} up to \SI{1000}{\volt}.
At every bias point $10^6$ trigger hits were recorded. Due to time constraints, not all combinations of voltages and designs could be measured.
The raw data was clustered and analysed with \textit{fei4Analyzer} \cite{fei4Analyzer_web}.
%\footnote{\url{https://github.com/terzo/fei4Analyzer}}
A Landau-Gauss convolution was fitted to the distribution of ToT values. The most probable value (MPV) of this function is plotted against the applied bias voltage in \autoref{fig:mpv_vs_bias}.

For all structures the MPV increases with the bias voltage.
The highest MPV at each voltage is measured at the structure with design V5.
As for the leakage current, good agreement between the results of designs 05 and V0 is found.
For all structures with modified pixel designs, except for V2, the collected charge is increased.
%While the increment is about \SI{19}{\percent} (\SI{32}{\percent}) for V3 (V1, V4, V6), it is more than \SI{100}{\percent} for V5.
The average increment at \SI{600}{\volt}, \SI{800}{\volt} and \SI{1000}{\volt} with respect to the IBL design is about \SI{19}{\percent} for V3, about \SI{32}{\percent} for V1, V4 and V6 and more than \SI{100}{\percent} for V5.

\section{Testbeam measurements}
Measurements were performed with a pion beam of \SI{120}{GeV} at CERN-SPS beamline H6. High tracking resolution of about \SI{5}{\micro\meter} in x and y was provided by six \mbox{MIMOSA26} sensors of ACONITE, an EUDET-type telescope \cite{Jansen2016,Bisanz_2017}. 
An unirradiated FE-I4 planar pixel module was used as reference plane.
All modules were placed in an isolated box, which was cooled with a regulated chiller and flushed with nitrogen. 
An on-sensor temperature of about \SI{-28}{\celsius} was determined for the device under test (DUT), which was then tuned to a threshold of \SI{3200}{e} and a ToT-response of \num{6} at a reference charge of \SI{20}{ke}.

\begin{figure}[htbp]
    \centering % \begin{center}/\end{center} takes some additional vertical space
    \includegraphics[width=1.\linewidth]{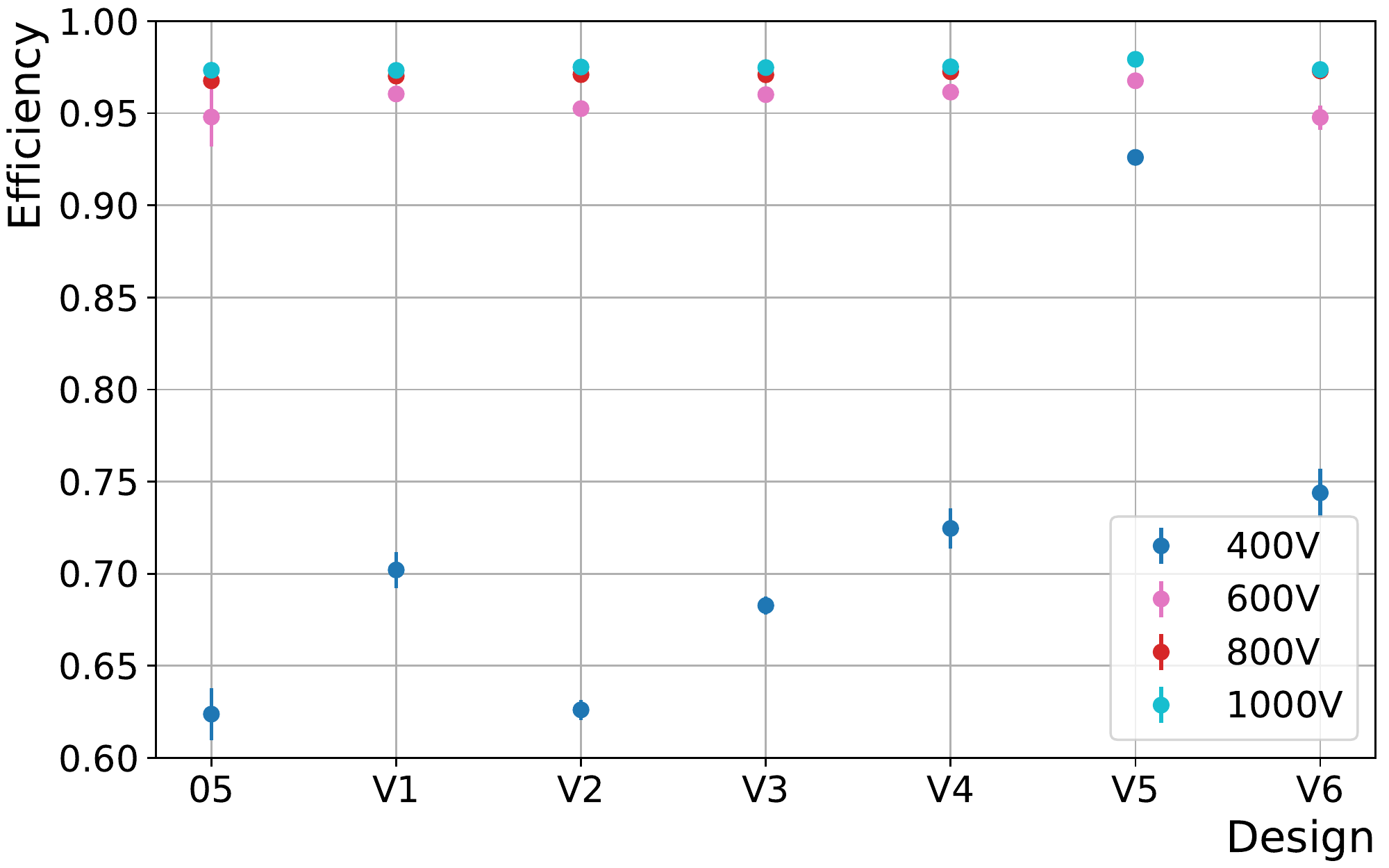}
    \caption{\label{fig:eff_vs_design} Efficiencies for the individual designs, at different bias voltages. Due to missing overlap between the telescope planes and the DUT, no data is available for the design V0. The error bars represent the standard error of the mean. Some are too small to be visible.}
\end{figure}

Track reconstruction was performed with \textit{EUTelescope} \cite{EUTelescope_web,Bisanz_2017}.
%\footnote{\url{http://eutelescope.web.cern.ch/}}
Due to different readout times of the telescope planes and the DUTs, only tracks given by the telescope which match to hits in the reference plane are considered in the number of tracks $n_\text{tracks}$. Each track which can be matched to a hit in the DUT is considered in the number of hits $n_\text{hits}$.
The efficiency $\varepsilon$ is defined as
\begin{linenomath*}
\begin{equation}
	\varepsilon = \frac{n_\text{hits}}{n_\text{tracks}},
\end{equation}
\end{linenomath*}
its relative error $\sigma_\varepsilon$ is defined as
\begin{linenomath*}
\begin{equation}
	\sigma_\varepsilon = \sqrt{\frac{\varepsilon \cdot (1 - \varepsilon)}{n_\text{tracks}}}.
\end{equation}
\end{linenomath*}
%The recorded data were divided into runs with roughly \SI{210}{k} events each. 
%The efficiency and its error were calculated for each run and pixel design separately, using the \textit{EfficiencyVsGeometry} analysis of \textit{TBmon2}\cite{TBmon2_web,Bisanz_2017}.
%\footnote{\url{https://bitbucket.org/TBmon2/tbmon2}}
%A weighted mean was then determined for all runs taken under the same conditions.
The data was recorded in runs, each containing roughly \SI{210}{k} events.
As the number of events per design strongly depends on the beam position, the \textit{EfficiencyVsGeometry} analysis of \textit{TBmon2} \cite{TBmon2_web,Bisanz_2017} was used to calculate the efficiency and its error separately for each pixel design and each run.
For each design, a mean weighted by the number of events was determined for all runs taken under the same conditions.
The efficiencies for the individual designs at different bias voltages are plotted in \autoref{fig:eff_vs_design}. 
%Due to positioning reasons in the setup, no data is available for the second IBL design V0.
Due to missing overlap between the telescope planes and the DUT, no data is available for the second IBL design V0.

For the higher voltage points saturation can be observed. The efficiencies of all designs are compatible with each other. 
At \SI{1000}{V}, the average efficiency is \SI{$(97.5 \pm 0.2)$}{\percent}, with design 05 showing the lowest value of \SI{$(97.3 \pm 0.3)$}{\percent}.
At \SI{800}{V}, the average efficiency is \SI{$(97.1 \pm 0.2)$}{\percent}, again with design 05 showing the lowest value of \SI{$(96.8 \pm 0.4)$}{\percent}.
The in-pixel efficiency maps at these voltages reveal that the inefficiencies are entirely in the region of the bias dot and the bias grid, independently from the design.

At \SI{600}{V} the first differences are visible and become more prominent at \SI{400}{V}.
The detailed data can be found in \autoref{tab:efficiencies_600V} and \autoref{tab:efficiencies_400V}.
\begin{table}[htb]
\centering
\caption{\label{tab:efficiencies_600V} Efficiencies for the individual designs, at \SI{600}{V}.}
\begin{tabular}{ccccc}
%\hline
\toprule
05 & \SI{$(94.8 \pm 1.6)$}{\percent} & & V4 & \SI{$(96.1 \pm 0.3)$}{\percent} \\ 
V1 & \SI{$(96.0 \pm 0.3)$}{\percent} & & V5 & \SI{$(96.8 \pm 0.2)$}{\percent} \\ 
V2 & \SI{$(95.2 \pm 0.3)$}{\percent} & & V6 & \SI{$(94.8 \pm 0.7)$}{\percent} \\ 
V3 & \SI{$(96.0 \pm 0.3)$}{\percent} & & &  \\  
%\hline
\bottomrule
\end{tabular}
%\end{table}
%\begin{table}[ht]
\centering
\caption{\label{tab:efficiencies_400V} Efficiencies for the individual designs, at \SI{400}{V}.}
\begin{tabular}{ccccc}
%\hline
\toprule
05 & \SI{$(62.4 \pm 1.4)$}{\percent} & & V4 & \SI{$(72.5 \pm 1.1)$}{\percent} \\ 
V1 & \SI{$(70.2 \pm 1.0)$}{\percent} & & V5 & \SI{$(92.6 \pm 0.2)$}{\percent} \\ 
V2 & \SI{$(62.6 \pm 0.5)$}{\percent} & & V6 & \SI{$(74.4 \pm 1.3)$}{\percent} \\ 
V3 & \SI{$(68.3 \pm 0.5)$}{\percent} & & &  \\  
%\hline
\bottomrule
\end{tabular}
\end{table}

\begin{figure}[htbp]
	\centering
	\includegraphics[width=0.95\linewidth]{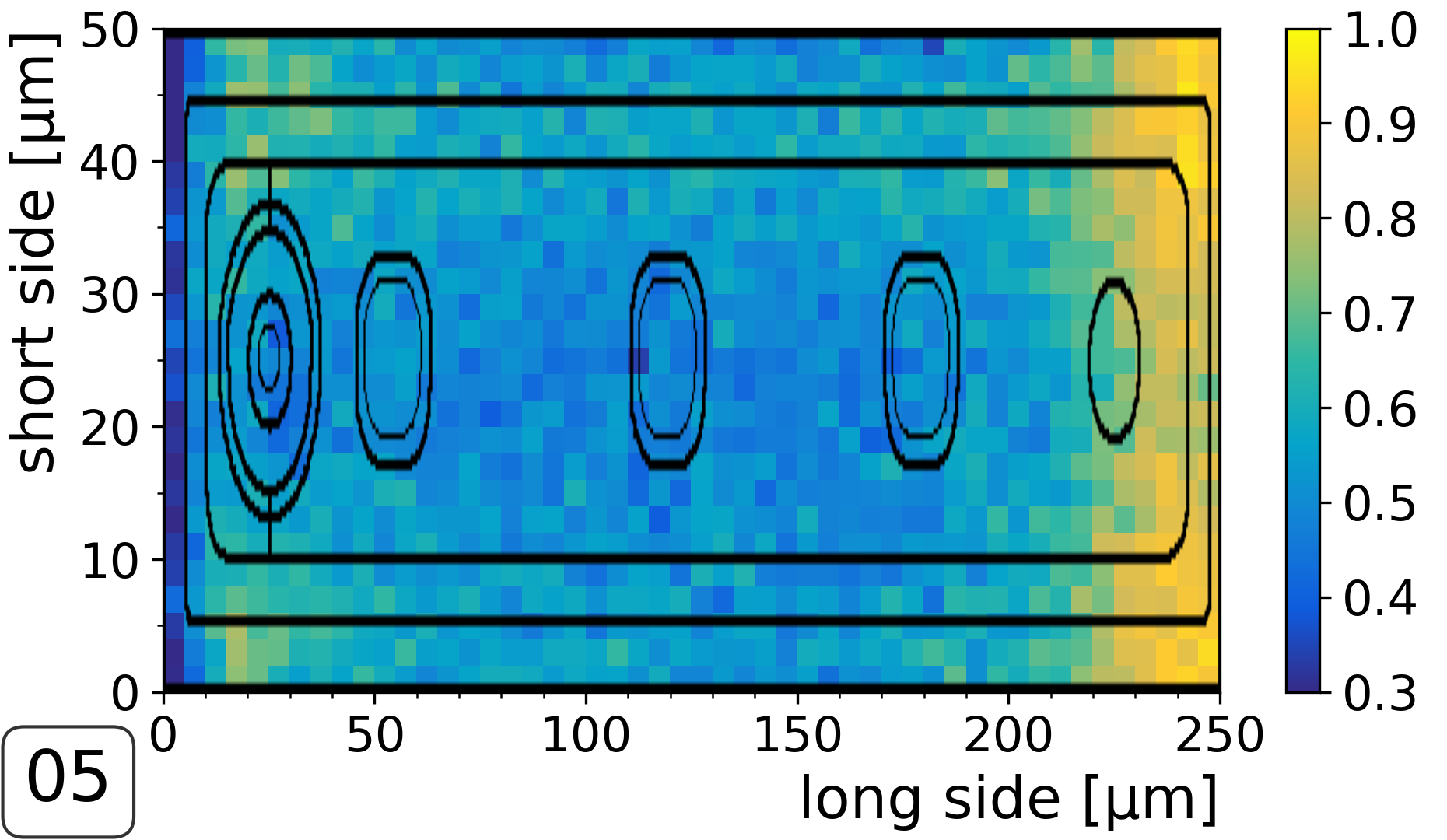}
	
	\vspace{20pt}
	\includegraphics[width=0.95\linewidth]{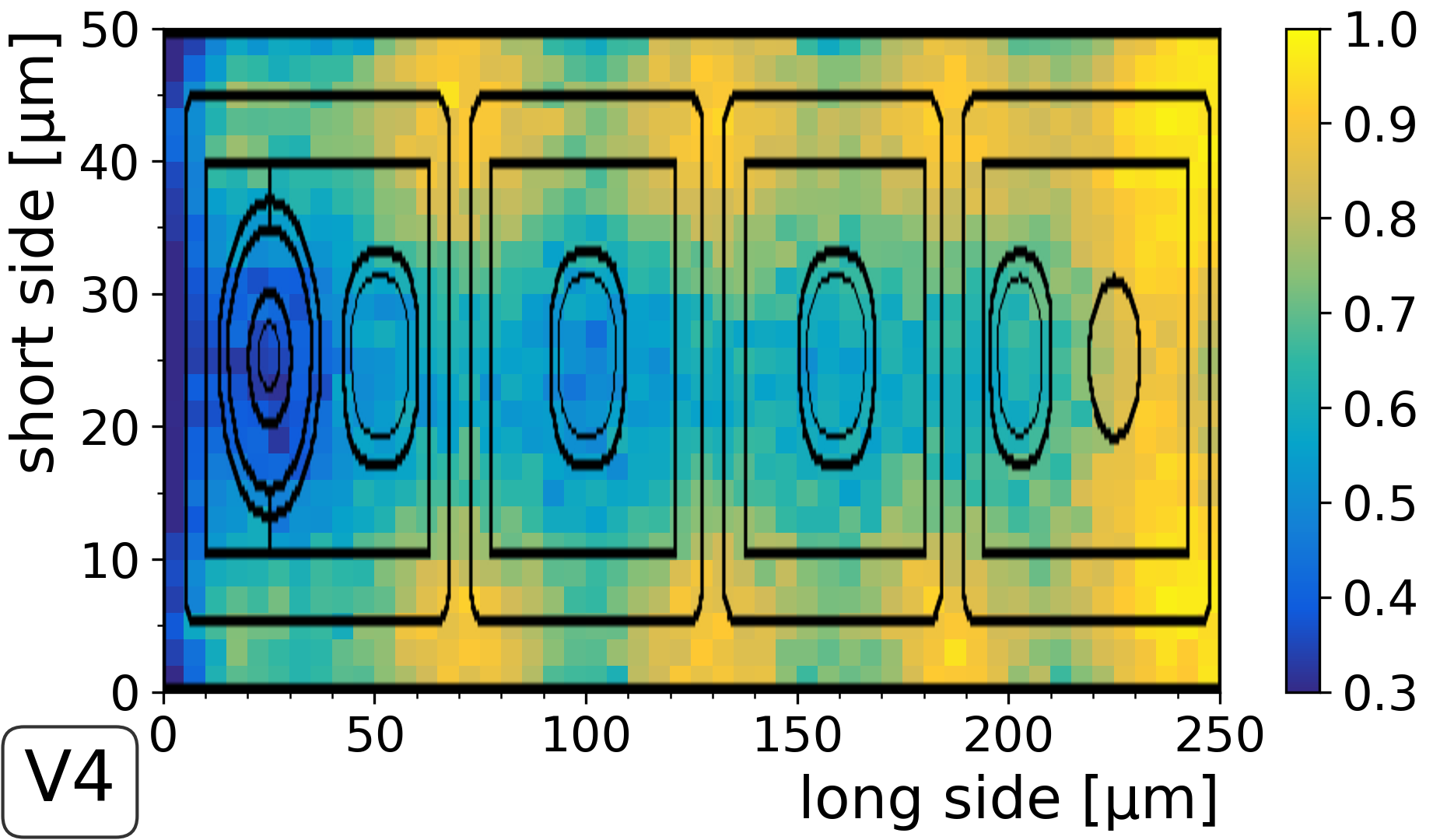}
	
	\vspace{20pt}
	\includegraphics[width=0.95\linewidth]{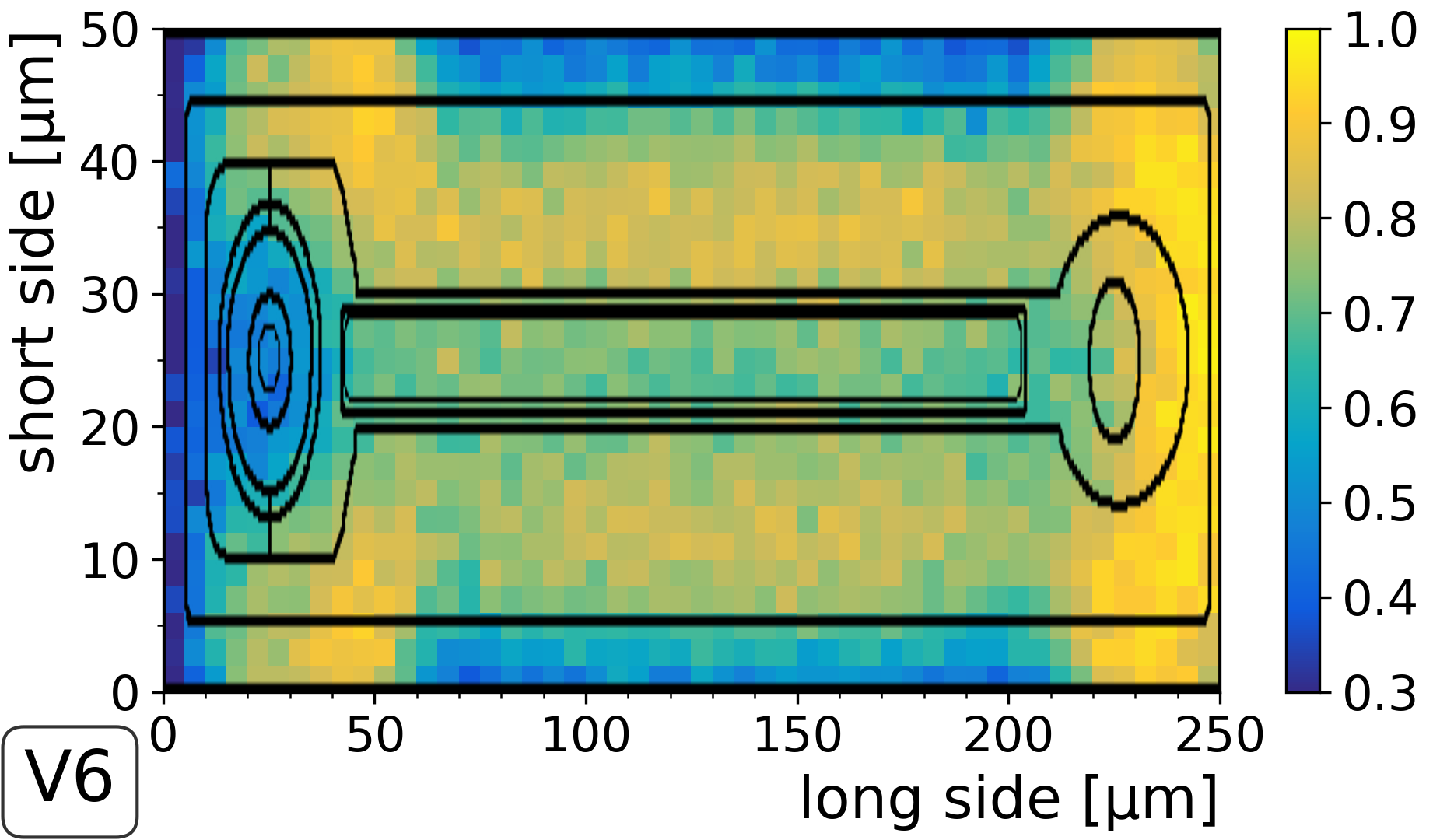}
	
	\vspace{20pt}
	\includegraphics[width=0.95\linewidth]{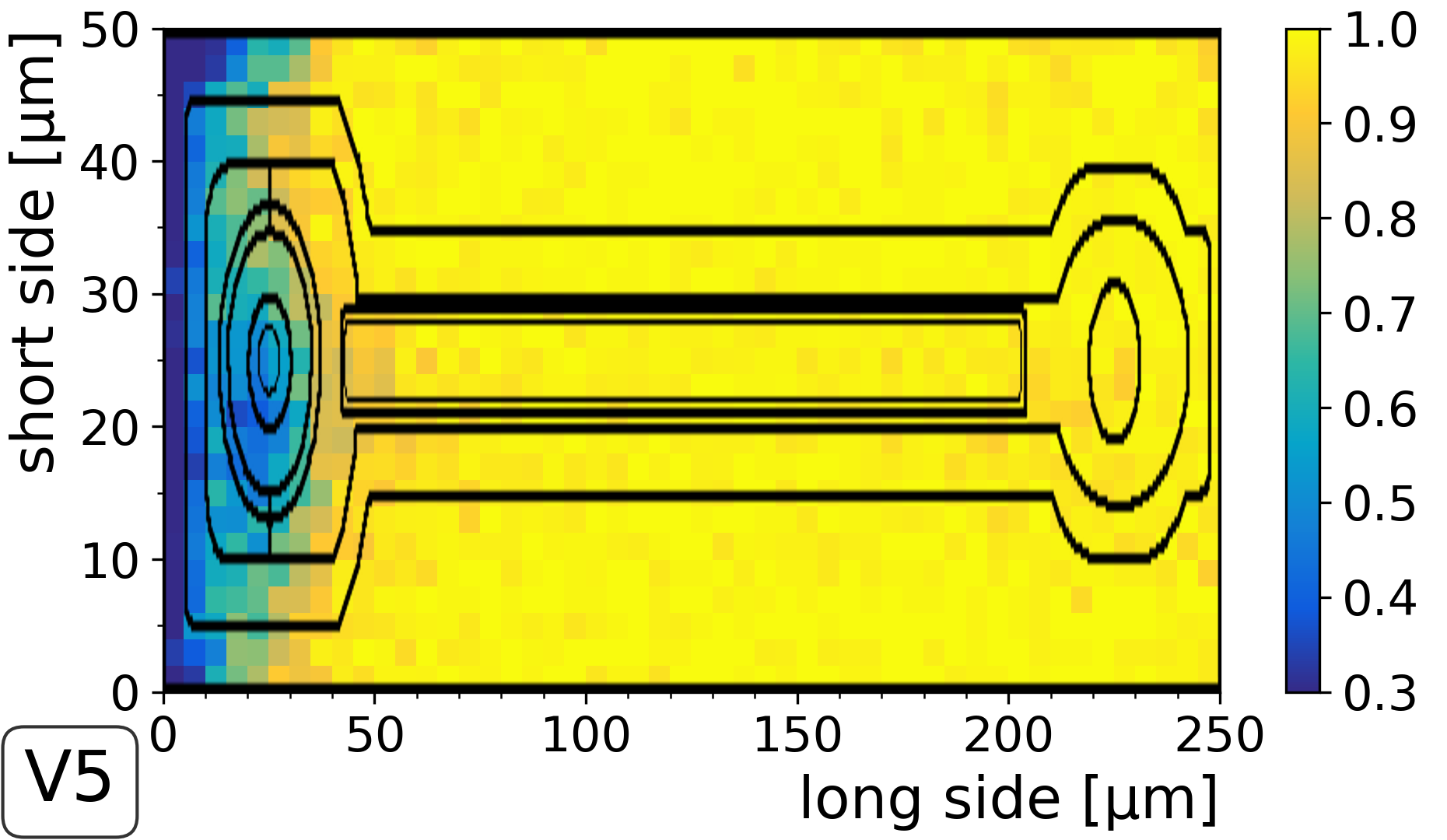}
    \caption{\label{fig:in-pixel_efficiencies} In-pixel efficiency maps of design 05, V4, V6 and V5, at \SI{400}{\volt}, overlayed with their schematic layouts. The bias grid and the bias dot is on the left, the bump bond connection is on the right.}
\end{figure}

The in-pixel efficiency maps in \autoref{fig:in-pixel_efficiencies} reveal where the differences come from.
The efficiency in the IBL design 05 is uniformly distributed, only some edge effects are visible.
For V4 and V6 efficiency hot spots are visible, which can be clearly connected to the design.
The main area of V5 is highly efficient. Even at low voltages, this design suffers only from inefficiencies caused by the bias dot and the bias grid.

\section{Summary and Outlook}

Several sensors with \emph{REINER} pixel implantations have been characterized in the laboratory and at testbeams.
On a module irradiated with neutrons to \SI{$5 \times 10^{15}$}{\neq}, the \emph{REINER} designs perform similar or better than the IBL standard design in terms of charge collection and tracking efficiency, at the cost of a slightly increased leakage current. 
%The same results can be achieved with reduced bias voltage.
%Comparing design V5 to the IBL design, the same hit efficiency is archived while the bias voltage is reduced by \SI{200}{\volt}, leading to a sensor power reduction of at least \SI{12}{\percent}.
%Comparing design V5 to the IBL design, V5 reaches at \SI{800}{\volt} (\SI{600}{\volt}) the same efficiency of \SI{97.4}{\percent} (\SI{96.8}{\percent}) as 05 at \SI{1000}{\volt} (\SI{800}{\volt}), leading to a sensor power reduction of \SI{12}{\percent} (\SI{43}{\percent}).
Comparing design V5 to the IBL design 05, V5 reaches at \SI{800}{\volt} the same efficiency of \SI{97.4}{\percent} as 05 at \SI{1000}{\volt}, leading to a sensor power reduction of \SI{12}{\percent}.
A larger power reduction can be achieved at lower bias voltages.
V5 reaches at \SI{600}{\volt} the same efficiency of \SI{96.8}{\percent} as 05 at \SI{800}{\volt} with the sensor power reduced by \SI{43}{\percent}.
 
Designs with larger areas of high-dosed p-spray (V1, V4, V5) perform better than designs without (V2, V3, V6), while the effect of the n$^+$ implantation shape seems minor.
The amplification seems to be induced by the additional p-n junction.
It is probably a gain effect similar to that already used in avalanche diodes and LGADs \cite{PELLEGRINI2014}.

%New designs based on this results were created, sensor delivery expected this month
To continue the studies, the same measurements are currently performed on a second \emph{REINER} pixel module which was irradiated with protons to about \SI{$6 \times 10^{15}$}{\neq}.
To investigate whether these results are easily transferable, all designs were also adopted to n$^+$-in-p and a wafer production was submitted.

\section{Acknowledgements}

The possibility to irradiate detector samples at the Sandia Annular Core Research Reactor is kindly acknowledged, 
%especially the organization of the campaign by M. Hoeferkamp and S. Seidel.
especially the help of M. Hoeferkamp and S. Seidel.

Special thanks go to all participants of the ATLAS ITk pixel testbeam campaigns, especially those who develop and maintain the corresponding hardware and software.

The work presented here is carried out within the framework of Forschungsschwerpunkt FSP103 and supported by
the Bundesministerium f\"ur Bildung und Forschung BMBF under grants 05H15PECAA and 05H15PECA9.

\section*{References}

\bibliography{bib-hstd11}

\begin{thebibliography}{10}
\expandafter\ifx\csname url\endcsname\relax
  \def\url#1{\texttt{#1}}\fi
\expandafter\ifx\csname urlprefix\endcsname\relax\def\urlprefix{URL }\fi
\expandafter\ifx\csname href\endcsname\relax
  \def\href#1#2{#2} \def\path#1{#1}\fi

\bibitem{ATLAS:Experiment}
{ATLAS Collaboration}, {The ATLAS Experiment at the CERN Large Hadron
  Collider}, JINST 3~(08) (2008) S08003.
\newblock \href {http://dx.doi.org/10.1088/1748-0221/3/08/S08003}
  {\path{doi:10.1088/1748-0221/3/08/S08003}}.

\bibitem{ATLAS:IBL_TDR}
{ATLAS Collaboration}, \href{http://cds.cern.ch/record/1291633/}{{ATLAS
  Insertable B-Layer Technical Design Report}}, Tech. Rep. CERN-LHCC-2010-013,
  CERN (2010).
\newline\urlprefix\url{http://cds.cern.ch/record/1291633/}

\bibitem{ATLAS:IBL_TDR2}
{ATLAS Collaboration}, \href{http://cds.cern.ch/record/1451888}{{ATLAS
  Insertable B-Layer Technical Design Report Addendum}}, Tech. Rep.
  CERN-LHCC-2012-009, CERN, addendum to CERN-LHCC-2010-013, CERN (2012).
\newline\urlprefix\url{http://cds.cern.ch/record/1451888}

\bibitem{1748-0221-7-02-C02051}
S.~Altenheiner, et~al., {Planar slim-edge pixel sensors for the ATLAS
  upgrades}, JINST 7~(02) (2012) C02051.
\newblock \href {http://dx.doi.org/10.1088/1748-0221/7/02/C02051}
  {\path{doi:10.1088/1748-0221/7/02/C02051}}.

\bibitem{GARCIASCIVERES2011S155}
M.~Garcia-Sciveres, et~al., {The FE-I4 pixel readout integrated circuit}, Nucl.
  Instr. Meth. Phys. Res. A 636 (2011) 155 -- 159.
\newblock \href {http://dx.doi.org/10.1016/j.nima.2010.04.101}
  {\path{doi:10.1016/j.nima.2010.04.101}}.

\bibitem{Casse2010}
G.~Casse, et~al., {Enhanced efficiency of segmented silicon detectors of
  different thicknesses after proton irradiations up to $1\times10^{16}$\neq},
  Nucl. Instr. Meth. Phys. Res. A 624 (2010) 401 -- 404.
\newblock \href {http://dx.doi.org/10.1016/j.nima.2010.02.134}
  {\path{doi:10.1016/j.nima.2010.02.134}}.

\bibitem{Kramberger2010}
G.~Kramberger, et~al., {Investigation of Irradiated Silicon Detectors by
  Edge-TCT}, IEEE Transactions on Nuclear Science 57 (2010) 2294--2302.
\newblock \href {http://dx.doi.org/10.1109/TNS.2010.2051957}
  {\path{doi:10.1109/TNS.2010.2051957}}.

\bibitem{Wittig2013}
T.~Wittig, {Slim Edge Studies, Design and Quality Control of Planar ATLAS IBL
  Pixel Sensors}, Ph.D. thesis, TU Dortmund (2013).
\newblock \href {http://dx.doi.org/10.17877/DE290R-5402}
  {\path{doi:10.17877/DE290R-5402}}.

\bibitem{Weingarten12}
J.~Weingarten, et~al., {Planar pixel sensors for the ATLAS upgrade: beam tests
  results}, JINST 7~(10) (2012) P10028.
\newblock \href {http://dx.doi.org/10.1088/1748-0221/7/10/P10028}
  {\path{doi:10.1088/1748-0221/7/10/P10028}}.

\bibitem{ATLAS:IBL_Prototype_modules}
{ATLAS IBL Collaboration}, {Prototype ATLAS IBL modules using the FE-I4A
  front-end readout chip}, JINST 7~(11) (2012) P11010.
\newblock \href {http://dx.doi.org/10.1088/1748-0221/7/11/P11010}
  {\path{doi:10.1088/1748-0221/7/11/P11010}}.

\bibitem{acrr_web}
\href{http://www.sandia.gov/research/facilities/annular_core_research_reactor.html}{{Sandia
  ACRR website}}.
\newline\urlprefix\url{http://www.sandia.gov/research/facilities/annular_core_research_reactor.html}

\bibitem{KLINGENBERG2016}
R.~Klingenberg, et~al., {Power dissipation studies on planar n$^+$-in-n pixel
  sensors}, Nucl. Instr. Meth. Phys. Res. A 831 (2016) 105 -- 110.
\newblock \href {http://dx.doi.org/10.1016/j.nima.2016.04.045}
  {\path{doi:10.1016/j.nima.2016.04.045}}.

\bibitem{Gisen17}
A.~Gisen, et~al., {Planar n-in-n quad module prototypes for the ATLAS ITk
  upgrade at HL-LHC}, JINST 12~(12) (2017) C12032.
\newblock \href {http://dx.doi.org/10.1088/1748-0221/12/12/C12032}
  {\path{doi:10.1088/1748-0221/12/12/C12032}}.

\bibitem{fei4Analyzer_web}
\href{https://github.com/terzo/fei4Analyzer}{{fei4Analyzer website}}.
\newline\urlprefix\url{https://github.com/terzo/fei4Analyzer}

\bibitem{Jansen2016}
H.~Jansen, et~al., {Performance of the EUDET-type beam telescopes}, EPJ Tech.
  Instrum. 3 (2016) 7 -- 27.
\newblock \href {http://dx.doi.org/10.1140/epjti/s40485-016-0033-2}
  {\path{doi:10.1140/epjti/s40485-016-0033-2}}.

\bibitem{Bisanz_2017}
T.~Bisanz, {Test-beam activities and results for the ATLAS ITk pixel detector},
  JINST 12~(12) (2017) C12053.
\newblock \href {http://dx.doi.org/10.1088/1748-0221/12/12/C12053}
  {\path{doi:10.1088/1748-0221/12/12/C12053}}.

\bibitem{EUTelescope_web}
\href{http://eutelescope.web.cern.ch/}{{EUTelescope website}}.
\newline\urlprefix\url{http://eutelescope.web.cern.ch/}

\bibitem{TBmon2_web}
\href{https://bitbucket.org/TBmon2/tbmon2}{{TBmon2 website}}.
\newline\urlprefix\url{https://bitbucket.org/TBmon2/tbmon2}

\bibitem{PELLEGRINI2014}
G.~Pellegrini, et~al., {Technology developments and first measurements of Low
  Gain Avalanche Detectors (LGAD) for high energy physics applications}, Nucl.
  Instr. Meth. Phys. Res. A 765 (2014) 12 -- 16.
\newblock \href {http://dx.doi.org/10.1016/j.nima.2014.06.008}
  {\path{doi:10.1016/j.nima.2014.06.008}}.

\end{thebibliography}

\end{document}